\documentclass{article}
\usepackage{arxiv}

\usepackage[utf8]{inputenc} 
\usepackage[T1]{fontenc}    
\usepackage{hyperref}       
\usepackage{url}            
\usepackage{booktabs}       
\usepackage{amsfonts}       
\usepackage{nicefrac}       
\usepackage{microtype}      
\usepackage{graphicx}
\usepackage{doi}
\usepackage{amsmath}
\usepackage{indentfirst}
\usepackage{titlesec}
\usepackage{caption}
\usepackage{subcaption}
\usepackage{graphicx}
\usepackage{comment}
\usepackage[square,numbers]{natbib}
\usepackage[english]{babel}

\title{Estimating Time of Arrival of Vehicle Fleets with GCN based Traffic Prediction}

\date{\today}	
\hypersetup{
    colorlinks=true,
    linkcolor=black,
    citecolor=black,
    filecolor=black,
    urlcolor=black,
}

\author{
  Shivika Sharma, Nandini Mawane, Dhruthick Gowda M, Mayur Taware \\
  \textbf{Chetan Kumar, Yash Chandrashekhar Dixit, Rakshit Ramesh} \\ 
  IUDX Programme Unit \\
  Indian Institute of Science \\ 
  Bangalore
}

\renewcommand{\headeright}{ }
\renewcommand{\undertitle}{ }

\hypersetup{
pdftitle={Estimating time of arrival with GCN},
pdfsubject={q-bio.NC, q-bio.QM},
pdfauthor={David S.~Hippocampus, Elias D.~Striatum},
pdfkeywords={First keyword, Second keyword, More},
}
\titleformat{\paragraph}
{\normalfont\normalsize\bfseries}{\theparagraph}{1em}{}
\titlespacing*{\paragraph}
{0pt}{3.25ex plus 1ex minus .2ex}{1.5ex plus .2ex}

\begin{document}
\maketitle
\vspace{-5pt}
\begin{abstract}
This paper presents an effective framework for estimating time of arrival of vehicles (buses) in an Intelligent Transit Management System (ITMS) having sparse position updates. Our contributions towards this is firstly in implementing a constrained optimization based road linestring segmenting framework ensuring ideal segment lengths and segments with sufficient density of vehicle position measurements which will result in valid statistics for scenarios involving sparse position measurements. Over this we propose a comprehensive approach for predicting traffic delays and estimated time of vehicle arrival addressing both the spatial and temporal dependencies of traffic. The traffic delay model is built on top of the T-GCN architecture on which we optimally augment an adjacency matrix which models a complexly connected road network considering the degree of influence between road segments, enabling the traffic delay model to look beyond physical road connectivity in predicting traffic delays and therefore producing better estimates of arrival times to points along the designated route of the vehicles.
\end{abstract}


\section{Introduction}

Traffic prediction is an integral part of traffic management system that helps in planning and controlling traffic by analysing trends. It provides a scientific basis for transit managers to sense traffic jams and prevent congestion. This has several applications that play a vital role in enhancing urban travel efficiency, including notifying passengers of real-time transit schedules, predicting traffic anomalies, determining real-time optimal transit routes, and identifying anomalous traffic hotspots. 
Traffic is also highly correlated amongst different roads owing to the connected nature of the network. Therefore it is important to have a model accounting for the same in Estimating the arrival times of public transit vehicles that operate in this network.

Generally \textit{Intelligent Transit Management Systems} in developing nations don't have vehicles fitted with state of the art position acquisition and transmission systems. A reasonable position update rate for vehicles as seen empirically by us across various cities is anywhere between 1 to 3 minutes. This is owing to various reasons such as low-cost hardware transmitting over 2G/GPRS radios and the general loss of signal receptivity in diverse urban environments. 
Even considering the lower rate of 1 minute position updates can cause significant uncertainties in estimating time of arrival considering the sporadic outbreaks of unexpected traffic snarls in developing nations like India.
Therefore it becomes imperative for a consumer facing time of arrival application to incorporate for these various anomalies.

We propose in this paper a framework to solve the problems of 
\begin{itemize}
    \item Finding appropriate spatial "bins" over which reasonably accurate statistical inferences such as average speeds and dwelling time can be estimated
    \item A Deep Learning based framework which can incorporate and account for sporadic, interconnected and rippling traffic scenarios
    \item A realtime forecasting strategy which takes into account the above and provides more frequent position updates for end user consumption
\end{itemize}

Towards realizing the above, we have developed a constrained optimization based approach to find these ideal spatial "bins" which ensures an ideal message density per bin/segment which is learned iteratively over long periods of historic data. This "segmentation" of a road line-string will also need to ensure that the generated segments are short enough so that the variability of traffic statistics such as average speed over its length is not beyond desired thresholds.

We then use these spatial bins / segments in a comprehensive approach for predicting segment-wise traffic delays. The proposed traffic delay model is built on top of the T-GCN architecture which previously was proposed for modelling traffic on a bus stop-to-stop basis. Therefore the required adjacency matrix incorporated only the connectivity of such stops. We augment this adjacency matrix with spatial correlations observed in long periods of historic data with which we believe complex road dynamics such as merging roads, far off unconnected bottlenecks at intersections and similarities in traffic patterns of roads unconnected to each other but having similar land use characteristics such as the presence of schools can also be captured.

The traffic delay model then serves as a dynamic input to standard Estimated Time of Arrival techniques in giving fine-grained spatio-temporal estimates of the vehicles position therefore accounting for the sporadic traffic outbreaks and sparsity in position estimates. We then show the performance of the framework in a real world scenario and provide reasonably accurate estimated time of arrivals for consumer applications.

\section{Related Works}
Several works have been proposed for segmenting large-scale urban road networks and predicting traffic delays and time of vehicle arrivals. Ghadi et. al. \citep{segmentationmethods} illustrate that the K-means clustering method for road segmentation performs better when compared against HSM method, constant AADT, constant length segmentation and curves. Derrow-Pinion et. al. \citep{googlemap} at Google Maps used an encoder-processor-decoder Graph Neural Network model for its predictions of vehicle arrival. Hu et. al. \citep{uberdeepeta} at Uber proposed a transformer architecture to compute feature weights in its DeepETA work. Zhao et. al. \citep{tgcn} aptly combined spatial and temporal dependencies using a GCN in combination with GRU in their Temporal Graph Convolutional Network model. We present a comprehensive procedure of creating segments using the Douglas–Peucker algorithm \citep{douglaspeucker} on raw data and computing traffic delays and ETAs using a modified version of the T-GCN model.

\section{Methodology}
In this section we outline the procedure used for creation of optimal road segments and model for traffic delay prediction.
\subsection{Data Description and Pre-processing}
The data used in this work was extracted from the real-time public transit information collected by India Urban Data Exchange (IUDX) from the city of Surat, Gujarat. It includes both real-time positions of transit buses across the city and static information about the routes followed by the buses every day. The Surat Municipal Corporation has close to $500$ buses that together make approximately $2000$ trips in and around the city that contribute to this data. These buses transmit messages every $1-5$ minutes and are expected to follow one of $51$ unique routes that cover both the inner city and major parts of the city outskirts. The dataset consists of $5277$ road segments, created as described in \ref{sec:road-segmentation}. Given data availability, we model hourly traffic delay predictions for time slots between $8 AM$ to $8 PM$ for the transit network of the city of Surat.

\subsection{Road Segmentation}
\label{sec:road-segmentation}
Road segments are constructed by applying the Douglas–Peucker algorithm \citep{douglaspeucker}, parameters of which are calculated based on a cost function. Set operations are performed to obtain the final segments. 

\subsubsection{Douglas–Peucker algorithm for Route Segmentation}
The construction of segments of individual routes is done using the Douglas-Peucker algorithm \citep{douglaspeucker} on the route points that are obtained post map matching from the static routes’ dataset, accounting for noise in the form of inaccuracies in geospatial coordinates that do not always fall on the road traversed by a transmitting bus. The algorithm returns break points along the route that mark the ends of segments. Some regions report poor message density due to network issues, faulty trackers or bus schedules (that is, very few buses ply on some segments). It is important for individual road segments to receive a substantial number of messages transmitted from buses for metrics such as average speed per segment to be reasonably calculated. Moreover, important insights of traffic anomalies may be lost due to inefficiently long segments. The value of compression threshold, $\epsilon$, defines the granularity and the lengths of road segments along a given route. Computation of an optimal value of $\epsilon$ is necessary to ensure that segments that are not too message-deficient nor too long. To calculate this optimal value of $\epsilon$, a grid search optimization routine is performed on each route with a cost function that takes both the length and the number of messages per segment into account.

\subsubsection{Message Density Estimation}
To enforce a constraint based on number of messages transmitted by buses within segments, we define message density as,
\begin{equation}
    \text{Message Density (m)} = \frac{\text{Average Number of Messages per Trip $(n_s)$}}{\text{Segment Length (l)}}
\end{equation}

\begin{figure} \vspace{-10pt}
     \centering
     \begin{subfigure}[b]{0.4\textwidth} 
         \centering
         \includegraphics[height=5cm]{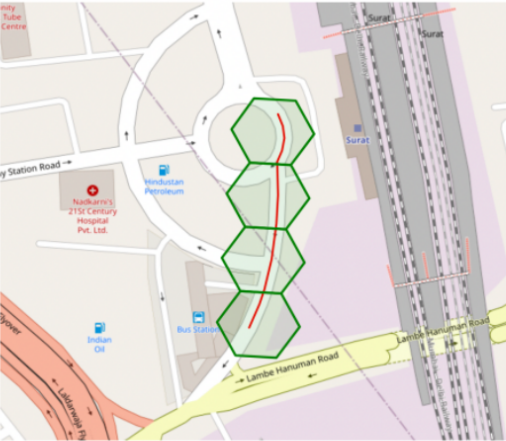}
         \caption{l = 1144, $n_s$ = 24, m = 0.021}
         \label{fig:messagedensity1}
     \end{subfigure} \quad
     \begin{subfigure}[b]{0.4\textwidth}
         \centering
         \includegraphics[height=5cm]{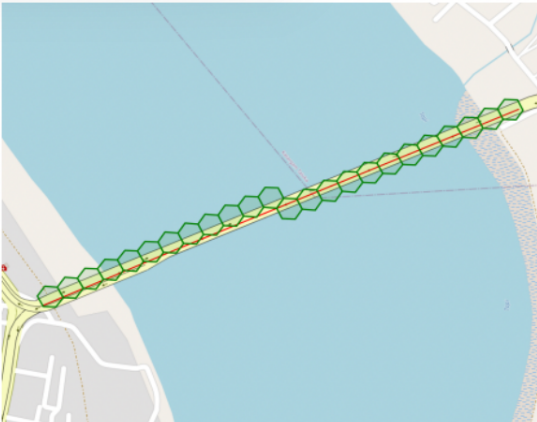}
         \caption{l = 154, $n_s$ = 24, m = 0.156}
         \label{fig:messagedensity2}
     \end{subfigure}
        \caption{Message Density Estimation using Uber’s H3}
        \label{fig:messagedensity} \vspace{-0.5pt}
\end{figure}

Upon estimation of average number of messages per trip using bus positions data and Uber’s H3 algorithm \citep{uberh3}, the bus positions are first binned into hexagons with an approximate edge length (\emph{25 metres} each). Then, a key-value dictionary \emph{(h: nh)} is constructed that maps each hexagon \emph{(h)} to the average number of messages transmitted by a trip through the hexagon \emph{(nh)}. This dictionary is used to estimate the average number of messages per trip within a segment as,
\begin{equation}
    n_s = \sum_{i=1} n_{h_i}
\end{equation}
where, $h_i$ are all the hexagons that fall on the line joining the segment’s two ends. \vspace{5pt} \\ 
For example, Figure \ref{fig:messagedensity} depicts two different segments and their respective hexagons. The longer segment, in Figure \ref{fig:messagedensity1}, over a long bridge across the river has the same average number of messages as the shorter segment, in Figure \ref{fig:messagedensity2}, that is part of a circle near a busy railway station, thus resulting in varying message densities between the two. Therefore, message density can be used as a measure to control the average message distribution across the segments of a route.

\subsubsection{Cost Function Definition and Optimization}
\begin{figure*} \vspace{-10pt}
  \begin{subfigure}{0.33\textwidth}
  \includegraphics[width=\textwidth]{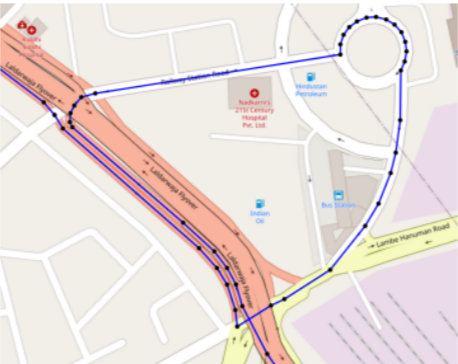}
  \caption{$\varepsilon = 0.00001$}
  \label{fig:merging-seg-1}
  \end{subfigure}
  \begin{subfigure}{0.33\textwidth}
  \includegraphics[width=\textwidth]{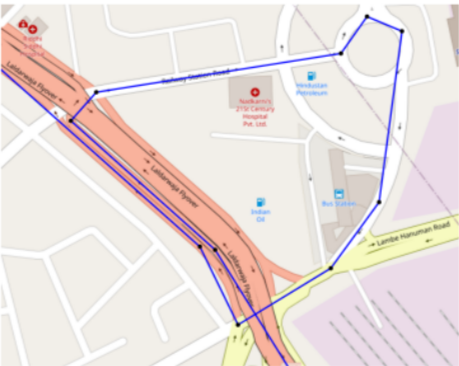}
  \caption{$\varepsilon = 0.0001$}
  \label{fig:route-seg-epsilon-2}
  \end{subfigure}
    \begin{subfigure}{0.33\textwidth}
  \includegraphics[width=\textwidth]{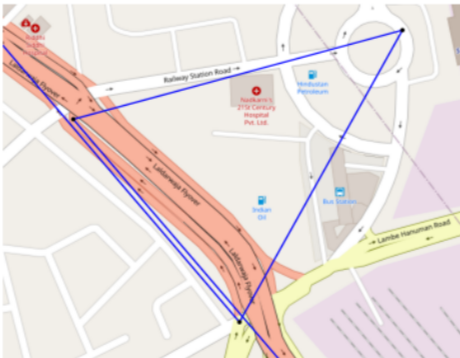}
  \caption{$\varepsilon = 0.001$}
  \label{fig:fig:route-seg-epsilon-3}
  \end{subfigure}\par\medskip
    \caption{Segments obtained for a section of Route 21D for different $\varepsilon$}
    \label{fig:route-seg-epsilon} \vspace{-10pt}
\end{figure*}

The value of the compression factor, $\epsilon$ in the Douglas-Peucker algorithm should result in segments that satisfy two constraints,
\begin{enumerate}
    \item Constraint on Messages: Number of transmitted messages within a segment must be substantial to perform meaningful analytical operations upon the final road segment network.
    \item Constraint on Length: The final segments cannot be unnecessarily long, in which case certain areas of the road network might be overlooked and relevant insights might be lost.
\end{enumerate}
For the above two constraints to be satisfied, an optimization routine using a simple grid search method is run on each route to calculate the optimal value for $\epsilon$ with a cost function defined as,
\begin{equation}
    \text{C($\epsilon$)} = \frac{\sigma\text{(m)}}{l_{avg}}
\end{equation}
\setlength{\parindent}{3em}
where, $\sigma(m)$ is the standard deviation of message densities, \newline
\indent $l_{avg}$ is the average segment length, \newline \indent  and  $10^{-4} < \epsilon < 10^{-3}$ \vspace{5pt} \newline 
The routine minimises the cost, which effectively reduces the variance in message density while simultaneously avoiding smaller segments by moderating the average segment length. The value of $\epsilon$ determines the number of points used to represent the entire route. As $\epsilon$ decreases, the number of segment break points obtained increases, as shown in Figure \ref{fig:route-seg-epsilon}, resulting in extremely small segments at circular junctions and roads that aren’t straight.

\setlength{\parindent}{0em}

\subsubsection{Extracting the Final Segments}
\begin{figure}
     \centering
     \begin{subfigure}[r]{0.45\textwidth}
         \centering
         \includegraphics[height=5cm]{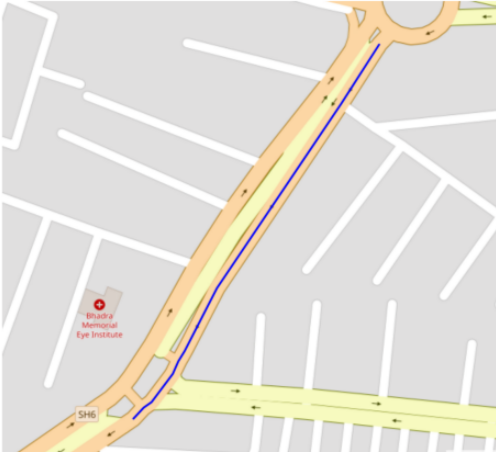}
         \caption{Segment from route 107U}
         \label{fig:seg-dilation-a}
     \end{subfigure} \hspace{-50pt}%
     \begin{subfigure}[l]{0.45\textwidth}
         \centering
         \includegraphics[height=5cm]{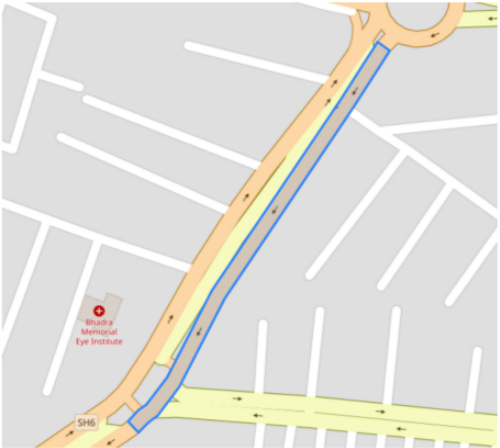}
         \caption{Dilation by 5 x 10$^{-4}$}
         \label{fig:seg-dilation-b}
     \end{subfigure}
        \caption{Dilating a segment}
        \label{fig:seg-dilation}
\end{figure}

\begin{figure*}
  \begin{subfigure}{0.33\textwidth}
  \includegraphics[width=\textwidth]{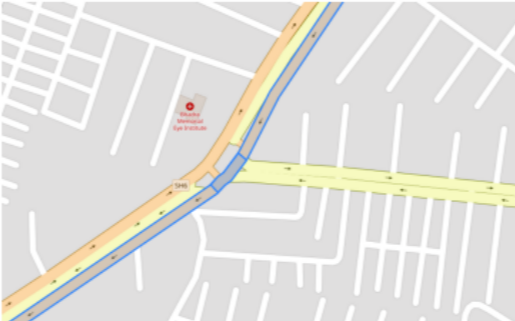}
  \caption{Two overlapping segments, s1 and s2}
  \label{fig:merging-seg-a}
  \end{subfigure}
  \begin{subfigure}{0.33\textwidth}
  \includegraphics[width=\textwidth]{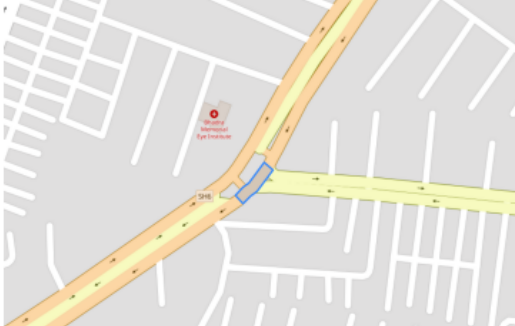}
  \caption{$s1 \cap s2$}
  \label{fig:merging-seg-b}
  \end{subfigure}
    \begin{subfigure}{0.33\textwidth}
  \includegraphics[width=\textwidth]{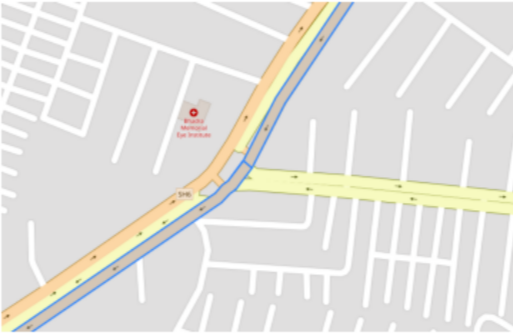}
  \caption{Final segments, s1 \textbackslash (s1 $\cap$ s2) and s2}
  \label{fig:merging-seg-c}
  \end{subfigure}\par\medskip
    \caption{Merging two overlapping segments s1 and s2}
    \label{fig:merging-seg}
\end{figure*}

Once the segments for all the routes are constructed, the next step is to merge the segments. Since bus routes in a city often intersect, this step ensures that route intersections are represented by the same segments. To deal with this, the segments obtained from individual routes are dilated into polygon segments. For example, Figure \ref{fig:seg-dilation} shows the result of dilation on a particular segment by a Euclidean distance of $5 \text{ x } 10^{-4}$. These polygon segments are then sorted by area in descending order before using set operations to merge them. Figure \ref{fig:merging-seg} shows how the overlapping region between two segments can be extracted through the intersection operation. The intersection operation along with differences can be used to extract the final segments that represent the entire transit road network of the city.

\begin{figure}
     \centering
     \begin{subfigure}[b]{0.4\textwidth}
         \centering
         \includegraphics[height=5cm]{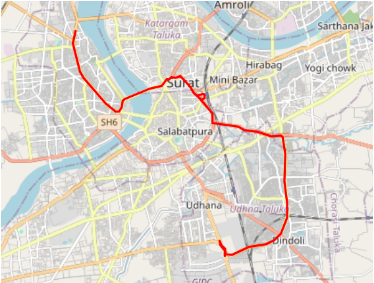}
         \caption{Route 21D}
         \label{fig:route-21d-a}
     \end{subfigure} \quad \hspace{-10pt}
     \begin{subfigure}[b]{0.4\textwidth}
         \centering
         \includegraphics[height=5cm]{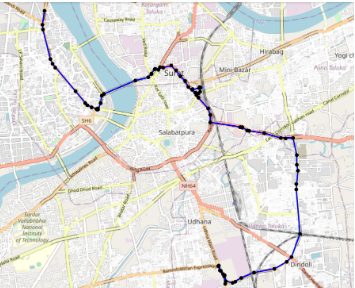}
         \caption{Route 21D segments with $\varepsilon$ = 0.0001}
         \label{fig:route-segmented-b}
     \end{subfigure}
        \caption{Using Douglas-Peucker algorithm to obtain segments}
        \label{fig:segmented-route}
\end{figure}

Figure \ref{fig:route-segmented-b} depicts the segments obtained for the route in Figure \ref{fig:route-21d-a} with $\epsilon=0.0001$. This route, composed of $1260$ points, is represented by $75$ segments.

\subsection{Traffic Delay Model}
In this section we discuss the model architecture (Figure \ref{fig:model-architecture}) and performance of the Traffic Delay Model that outputs dwell times for road segments which is time taken to travel the segment, for the next time slot given input time series data. 

\subsubsection{Model Architecture} 
\label{sec:model-architecture}
\begin{figure}
	\centering
	\includegraphics[width=0.5\textwidth]{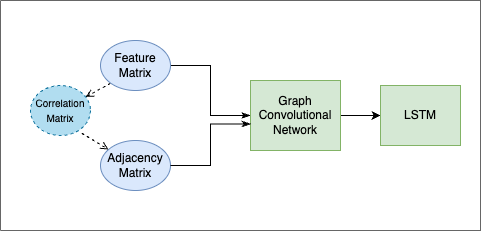}
	\caption{Traffic Delay Model}
	\label{fig:model-architecture}
\end{figure}

The Traffic Delay Model is a modified version of the Temporal Graph Convolutional Network (T-GCN) \citep{tgcn}. Using \emph{Graph Convolutional Network (GCN)} and \emph{Long Short Term Memory (LSTM)}, the model accommodates both spatial and temporal dependencies on segments created as described in \ref{sec:road-segmentation}. As shown in Figure \ref{fig:model-architecture}, the model architecture consists of 2 GCN layers with \emph{ReLU} activation followed by an LSTM. The inputs to the GCN layer are,
    \begin{enumerate}
        \item A feature matrix, $X\in R^{N x P}$ with $N$ segments, $P$ time slots
        and \vspace{-10pt} \newline \begin{center}
            $x_{ij}$ = dwell time for segment $i$ in time slot $j$
        \end{center} 
        \item An adjacency matrix of size $N$ x $N$ denoting segment connectivity
    \end{enumerate}
The output of the $GCN$ layer is then fed into an $LSTM$. Overall the model is given an input time series data of $H$ slots for $N$ segments for it to output the prediction for the $H+1^{th}$ time slot for all $N$ segments. Our main contribution is in the computation of the adjacency matrix that can optimally account for the traffic influence between segments.

\paragraph{Adjacency Matrix}\vspace{-5pt}
Given a representative road network graph $G=(V,E)$ where the nodes, $V$, denote segments and the undirected edges, $E$, represent influence between nodes, slight changes in the adjacency matrix result in changes in performance of the overall model. We present an optimal method (Equation \ref{eq:adjacency-calculation}) for the computation of these edge weights that goes beyond the traditional method of only looking at physical segment connectivity. In doing so, the graph convolution now considers the extent to which a segment influences another, hence also considering physically non-adjacent segments.
\begin{equation}
    \text{physical connectivity[i][j]}  = \text{\{1 if segments $i$ and $j$ are neighbours else 0\}}
\end{equation}
\begin{equation}
\label{eq:adjacency-calculation}
    \text{new adjacency[i][j]}  = \alpha * \text{physical connectivity[i][j]} + (1-\alpha)*\text{correlation[i][j]}; \alpha \in [0,1] \vspace{4pt}
\end{equation} 
The correlation matrix is computed using the feature matrix $X$ of the $T-GCN$ model which is then subjected to a threshold, $K$, for each $i, j \in V$ such that only positive and highly correlated segments are considered to reduce effects of noise in feature values.
\subsubsection{Traffic Delay Prediction Experiments}
\begin{figure}
	\centering
	\includegraphics[width=0.6\textwidth]{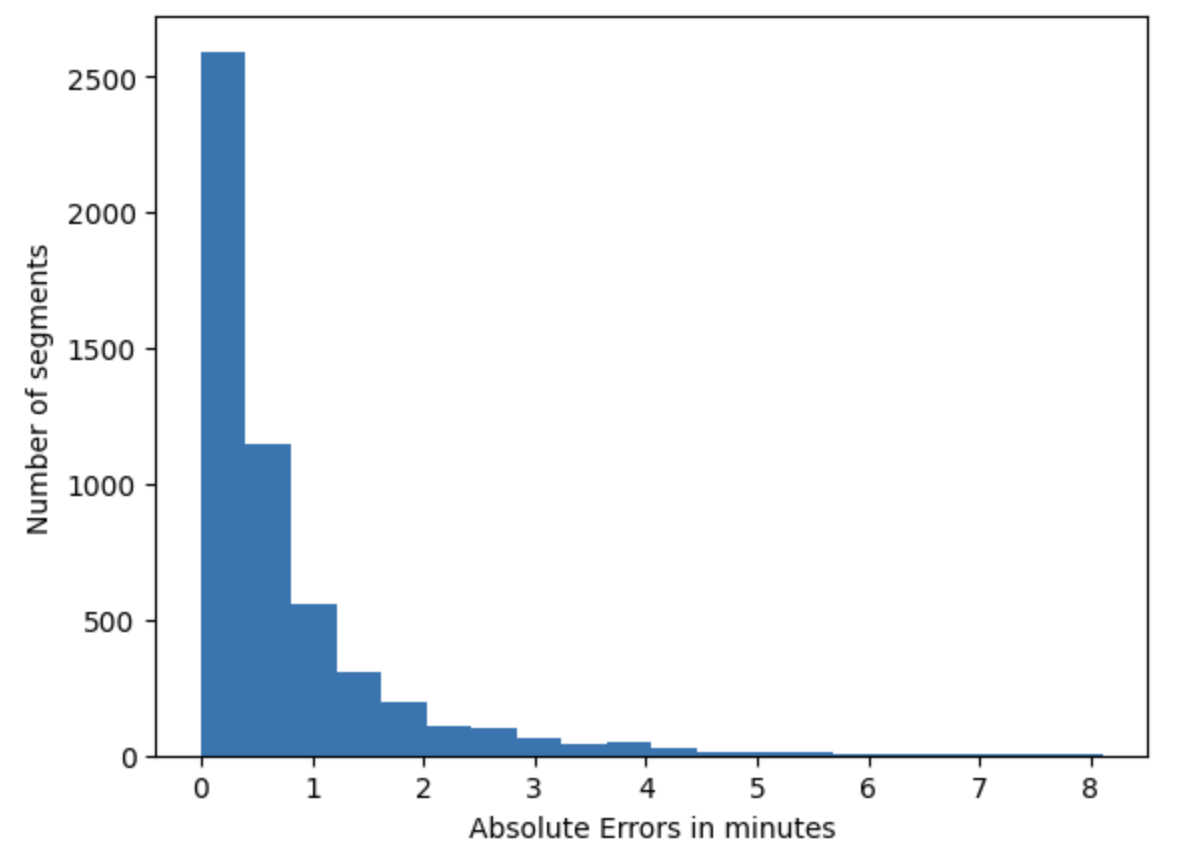}
	\caption{Distribution of absolute errors at $K$ = 0.65 and $\alpha$ = 0.85 averaged over all time slots}
	\label{fig:histogram}
\end{figure}

The feature matrix used for training is computed using three months of preprocessed segment dwell time data. The performance of the model is then analysed on the test data that is computed using preprocessed data of the fourth month. In our experiments, we use $H=4$. \vspace{2pt} \newline Through an exhaustive search, we find optimal values of hyperparameters $\alpha$ and $K$ described in \ref{sec:model-architecture} as $0.85$ and $0.65$ respectively. Table \ref{table:alpha-variation} reports RMSE in predicted traffic delays averaged over all considered time slots on varying $\alpha$. Further, we illustrate the distribution of errors for the found optimal values in Figure \ref{fig:histogram}. \vspace{-10pt} \newline

\begin{figure}
	\centering
	\includegraphics[width=0.8\textwidth]{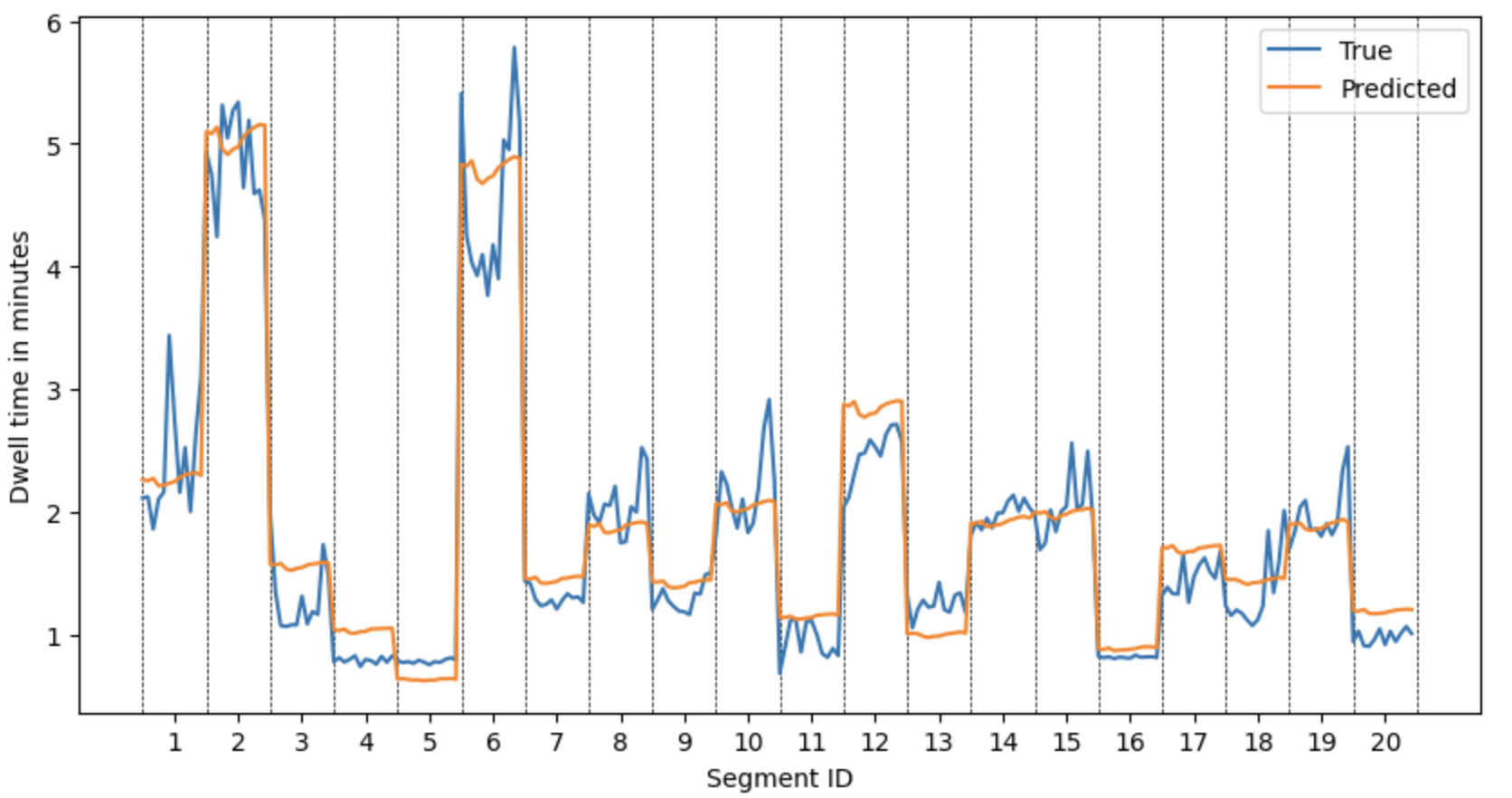}
	\caption{True and Predicted dwell times of 20 segments for different hour slots of the day}
	\label{fig:segment-dwelltimes}
\end{figure}

\begin{table} [h!]
	\centering
	\begin{tabular}{|c|c|}
		\hline \vspace{-8pt}\\
		$\alpha$     & RMSE in seconds  \\ [0.5ex]
		\hline \vspace{-8pt}\\
		1.00 (Traditional T-GCN) & 106.32 \\ [0.5ex]
		0.90 & 96.18 \\ [0.5ex]
            \textbf{0.85} & \textbf{95.82} \\ [0.5ex]
            0.80 & 96.78 \\ [0.5ex]
		\hline 
	\end{tabular}
        \captionsetup{skip=2pt}
        \caption{RMSE in traffic delay predictions on varying $\alpha$ with $K= 0.65$}
        \label{table:alpha-variation} \vspace{-10pt}
\end{table}
Figure \ref{fig:segment-dwelltimes} illustrates the true delays and delay predictions for segments over different time slots. 
\section{Application}
In this section, we present the methodology used to predict real-time estimated time of arrival using the traffic delay model described in \ref{sec:model-architecture}. The position of a bus is predicted after every $t$ seconds using the dwell time predictions of the traffic delay model for the $H+1^{th}$ time slot. These positions are updated using an Alpha-Beta filter \citep{alphabetafilter}, as real-time data is made available within the $H+1^{th}$ time slot. The ETA of the remaining trip is calculated by adding up the dwell times of the remaining segments. This method continuously updates predicted ETA and allows further refinement of predictions over the traffic model predictions. 
\begin{equation}
    \text{updated distance} =  \alpha'*\text{predicted distance} + (1-\alpha') * \text{measured distance} \vspace{-3pt}
\end{equation}

\begin{figure}
     \centering
     \begin{subfigure}[b]{0.49\textwidth}
         \centering
         \includegraphics[width=\textwidth]{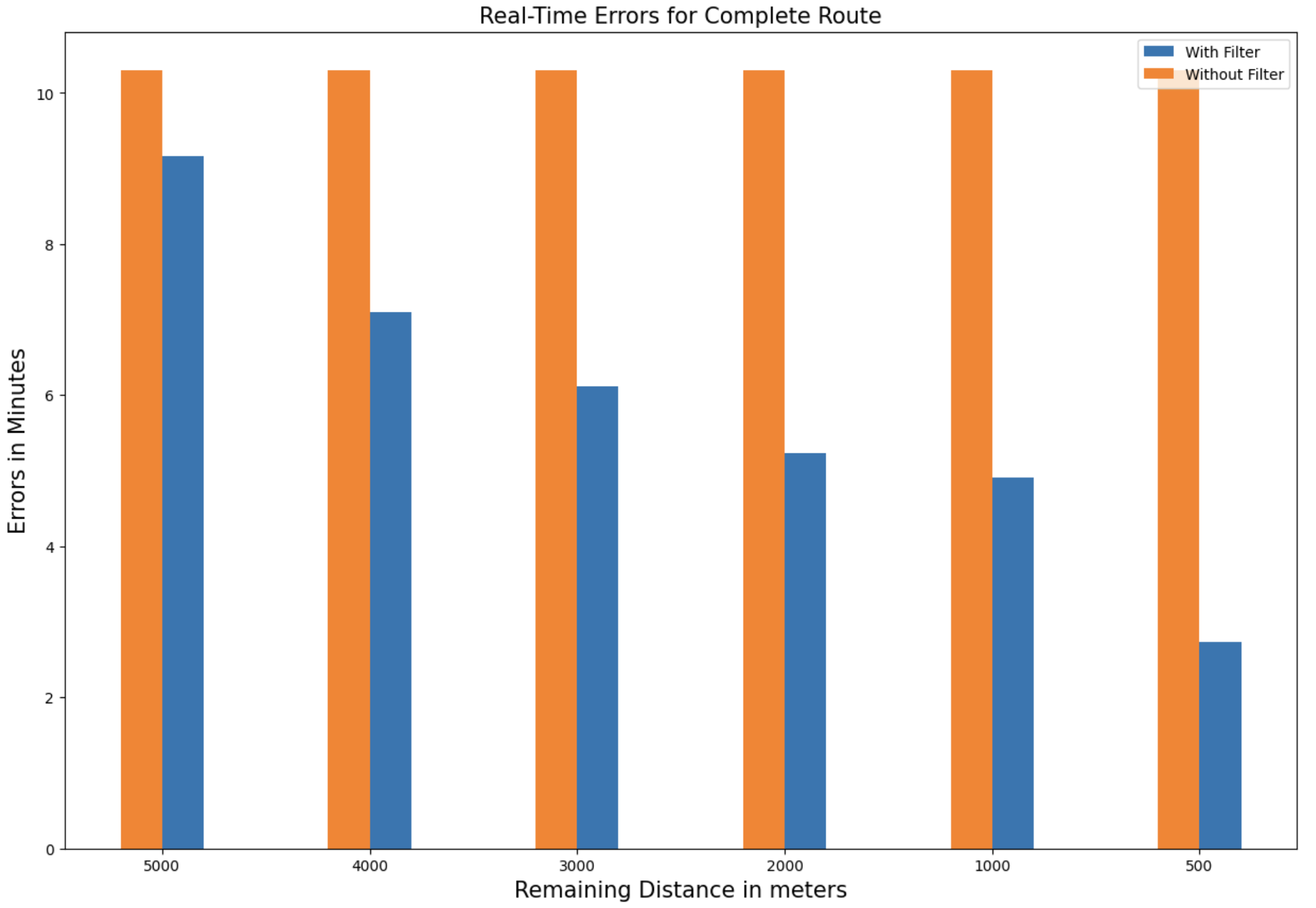}
         \caption{Mean Absolute Errors}
         \label{fig:application1}
     \end{subfigure} 
     \begin{subfigure}[b]{0.49\textwidth}
         \centering
         \includegraphics[width=\textwidth]{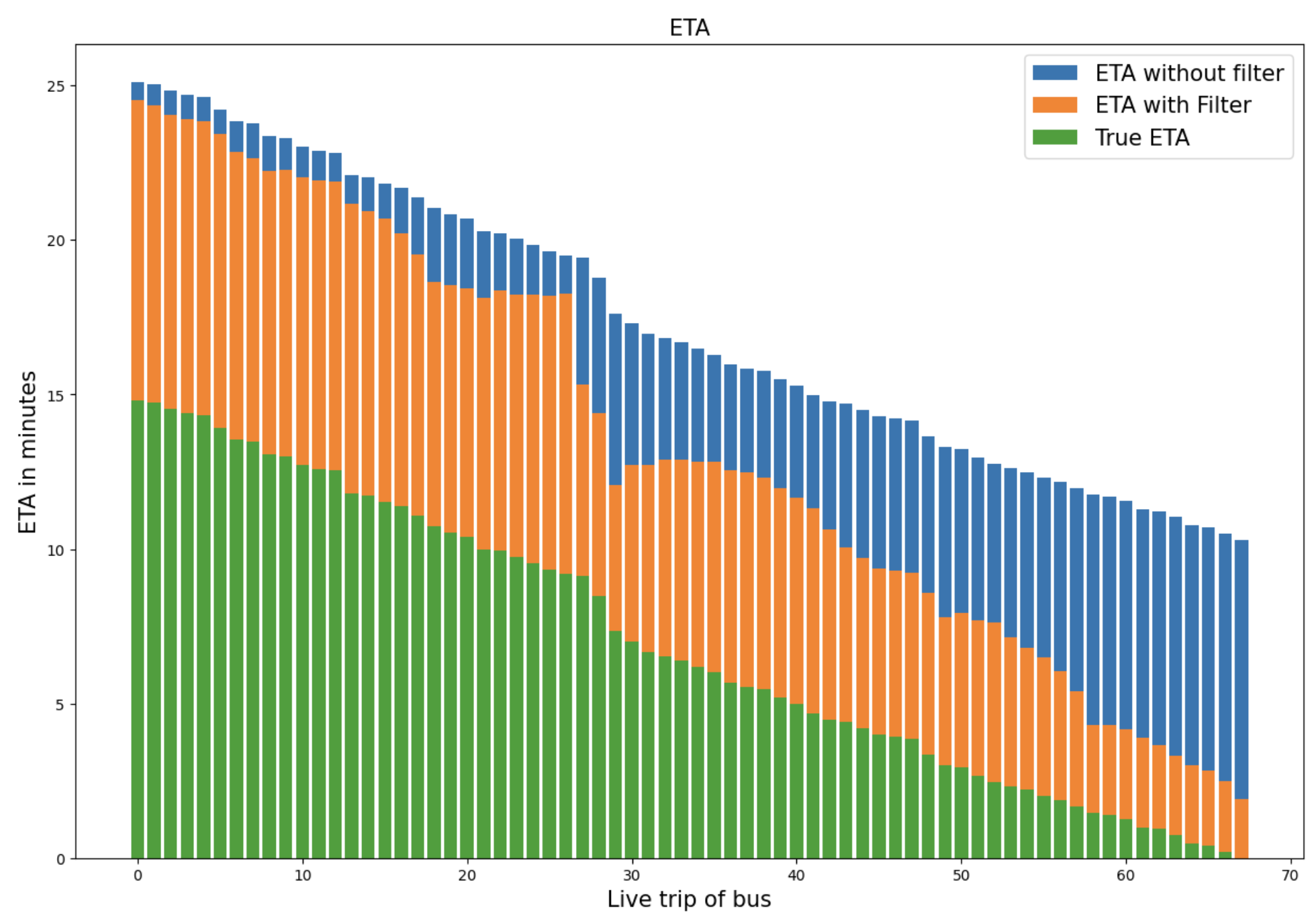}
         \caption{ETA prediction along a live trip}
         \label{fig:application2}
     \end{subfigure}
        \caption{ETA Prediction Results}
        \label{fig:application}
\end{figure}

Figure \ref{fig:application1} illustrates the mean absolute errors in predicted ETA upon varying the distance remaining for trip completion. Errors in the case of filtering (using $\alpha'$ = 0.9) decrease as distance decreases while errors without filtering remain constant.

\section{Conclusions}
This paper outlines an end-to-end procedure to predict traffic delays and estimated time of vehicle arrival using raw data. Road segments are efficiently created using the Douglas Peucker algorithm to find segments with optimal message densities and length. The model captures the influence of both other road segments and time, to predict the traffic delay of a segment. This model is further used to efficiently predict Estimated Time of Arrival using an Alpha-Beta filter on live data.

\bibliographystyle{abbrvnat}
\bibliography{references}  

\begin{thebibliography}{7}
\providecommand{\natexlab}[1]{#1}
\providecommand{\url}[1]{\texttt{#1}}
\expandafter\ifx\csname urlstyle\endcsname\relax
  \providecommand{\doi}[1]{doi: #1}\else
  \providecommand{\doi}{doi: \begingroup \urlstyle{rm}\Url}\fi

\bibitem[Austin Derrow-Pinion(2021)]{googlemap}
e.~a. Austin Derrow-Pinion, Jennifer~She.
\newblock Eta prediction with graph neural networks in google maps.
\newblock \emph{arXiv preprint arXiv:2108.11482}, 2021.
\newblock \doi{https://doi.org/10.48550/arXiv.2108.11482}.

\bibitem[Douglas and Peucker(1973)]{douglaspeucker}
D.~Douglas and T.~Peucker.
\newblock Algorithms for the reduction of the number of points required to represent a digitized line or its caricature.
\newblock \emph{Cartographica: The International Journal for Geographic Information and Geovisualization}, pages 112--122, 1973.
\newblock \doi{10.3138/FM57-6770-U75U-7727}.

\bibitem[et~al.(2019)]{segmentationmethods}
Q.~G. et~al.
\newblock Comparison of different road segmentation methods.
\newblock \emph{Promet - Traffic\&Transportation}, 2019.
\newblock \doi{https://doi.org/10.7307/ptt.v31i2.2937}.

\bibitem[Ling~Zhao(2018)]{tgcn}
Y.~S. e.~a. Ling~Zhao.
\newblock T-gcn: A temporal graph convolutionalnetwork for traffic prediction.
\newblock \emph{arXiv preprint arXiv:1811.05320}, 2018.
\newblock \doi{https://doi.org/10.48550/arXiv.1811.05320}.

\bibitem[Penoyer(1993)]{alphabetafilter}
R.~Penoyer.
\newblock The alpha-beta filter.
\newblock \emph{C Users Journal}, 1993.

\bibitem[Woźniak and Szymański(2018)]{uberh3}
S.~Woźniak and P.~Szymański.
\newblock hex2vec - context-aware embedding h3 hexagons with openstreetmap tags.
\newblock \emph{arXiv preprint arXiv:1804.09028}, 2018.
\newblock \doi{https://doi.org/10.48550/arXiv.2111.00970}.

\bibitem[Xinyu~Hu(2022)]{uberdeepeta}
T.~B. e.~a. Xinyu~Hu.
\newblock Deepreta: An eta post-processing system at scale.
\newblock \emph{arXiv preprint arXiv:2206.02127}, 2022.
\newblock \doi{https://doi.org/10.48550/arXiv.2206.02127}.

\end{thebibliography}
\end{document}